\documentclass[letterpaper, 10 pt, conference]{ieeeconf} 
\IEEEoverridecommandlockouts

\usepackage[letterpaper, left=0.75in, right=0.75in, bottom=0.8in, top=0.75in]{geometry}

\usepackage{cite}
\usepackage{lipsum}
\usepackage{booktabs}
\usepackage{amsmath,amssymb,amsfonts}
\usepackage{algorithmic}
\usepackage{graphicx}
\usepackage{textcomp}
\usepackage{xcolor}
\usepackage{graphicx}
\usepackage{subcaption}
\usepackage{multirow}
\usepackage{mwe}
\usepackage{pgfplots}
\usepackage{hyperref}
\def\BibTeX{{\rm B\kern-.05em{\sc i\kern-.025em b}\kern-.08em
    T\kern-.1667em\lower.7ex\hbox{E}\kern-.125emX}}

\begin{document}

\title{ODAS: Open embeddeD Audition System}

\author{Fran\c{c}ois Grondin, Dominic L\'etourneau, C\'edric Godin, Jean-Samuel Lauzon, Jonathan Vincent, \\Simon Michaud, Samuel Faucher, Fran\c{c}ois Michaud}

\maketitle


\begin{abstract}
Artificial audition aims at providing hearing capabilities to machines, computers and robots. Existing frameworks in robot audition offer interesting sound source localization, tracking and separation performance, although involve a significant amount of computations that limit their use on robots with embedded computing capabilities. This paper presents ODAS, the Open embeddeD Audition System framework, which includes strategies to reduce the computational load and perform robot audition tasks on low-cost embedded computing systems. It presents key features of ODAS, along with cases illustrating its uses in different robots and artificial audition applications.
\end{abstract}



\section{Introduction}

Similarly to artificial/computer vision, artificial/computer audition can be defined as the ability to provide hearing capabilities to machines, computers and robots. 
Vocal assistants on smart phones and smart speakers are now common, providing a vocal interface between people and devices \cite{hoy2018alexa}.
As for artificial vision, there are still many problems to resolve for endowing robots with adequate hearing capabilities, such as ego and non-stationary noise cancellation, mobile and distant speech and sound understanding \cite{ince2011assessment,deleforge2015phase,schmidt2018novel,rascon2018acoustic,shimada2019unsupervised}.

Open source software frameworks, such as OpenCV \cite{culjak2012brief} for vision and ROS \cite{Quigley2009} for robotics, greatly contribute in making these research fields evolve and progress, allowing the research community to share and mutually benefit from collective efforts.
In artificial audition, two main frameworks exist: 

\begin{itemize}
    \item HARK (Honda Research Institute Japan Audition for Robots with Kyoto University\footnote{\url{https://www.hark.jp/}}) provides multiple modules for sound source localization and separation \cite{nakadai2008open,nakadai2010design,nakadai2017development}.
    This framework is mostly built over the FlowDesigner software \cite{cote2004code}, and can also be interfaced with speech recognition tools such as Julius \cite{lee2009recent} and Kaldi \cite{povey2011kaldi,ravanelli2019pytorch}.
    HARK implements sound source localization in 2-D using variants of the Multiple Signal Classification (MUSIC) algorithm \cite{ishi2009evaluation,nakamura2009intelligent,nakamura2012real}.
    HARK also performs geometrically-constrained higher-order decorrelation-based source separation with adaptive step-size control\cite{okuno2015robot}.
    Though HARK supports numerous signal processing methods, it requires a significant amount of computing power (in part due to the numerous eigenvalue decompositions required by MUSIC), which makes it less suitable for use on low-cost embedding hardware.
    For instance, when using HARK with a drone equipped with a microphone array to perform sound source localization, the raw audio streams needs to be transferred on the ground to three laptops for processing \cite{nakadai2017development2}.

    \item ManyEars\footnote{\url{https://github.com/introlab/manyears}} is used with many robots to perform sound localization, tracking and separation \cite{grondin2013manyears}.
    Sound source localization in 3-D relies on Steered-Response Power with Phase Transform (SRP-PHAT), and tracking is done with particle filters \cite{valin2007robust}.
    ManyEars also implements the Geometric Sound Separation (GSS) algorithm to separate each target sound source \cite{parra2002geometric,valin2004enhanced}.
    This framework is coded in the C language to speed up computations, yet it remains challenging to run all algorithms simultaneously on low-cost embedding hardware such as a Digital Signal Processor (DSP) chip \cite{briere2008embedded}. 
\end{itemize}

Although both frameworks provide useful functionalities for robot audition tasks, they require a fair amount of computations.
There is therefore a need for a new framework providing artificial audition capabilities in real-time and running on low-cost hardware.
To this end, this paper presents ODAS\footnote{\url{http://odas.io}} (Open embeddeD Audition System), improving on the ManyEars framework by using strategies to optimize processing and performance.
The paper is organized as follows. 
Section \ref{sec:ODAS} presents ODAS' functionalities, followed by Section \ref{sec:lib} with configuration information of the ODAS library.
Section \ref{sec:applications} describes the use of the framework in different applications.

\section{ODAS}
\label{sec:ODAS}

As for ManyEars \cite{grondin2013manyears}, ODAS audio pipeline consists of a cascade of three main modules -- localization, tracking and separation -- plus a web interface for data vizualisation.
The ODAS framework also uses multiple I/O interfaces to get access to raw audio data from the microphones, and to return the potential directions of arrival (DOAs) generated by the localization module, the tracked DOAs produced by the tracking module and the separated audio streams.
ODAS is developed using the C programming language, and to maximize portability it only has one external dependency to the well-known third-party FFTW3 library (to perform efficient Fast Fourier Transform) \cite{frigo2005design}.

\begin{figure*}[!ht]
    \centering
    \includegraphics[width=\linewidth]{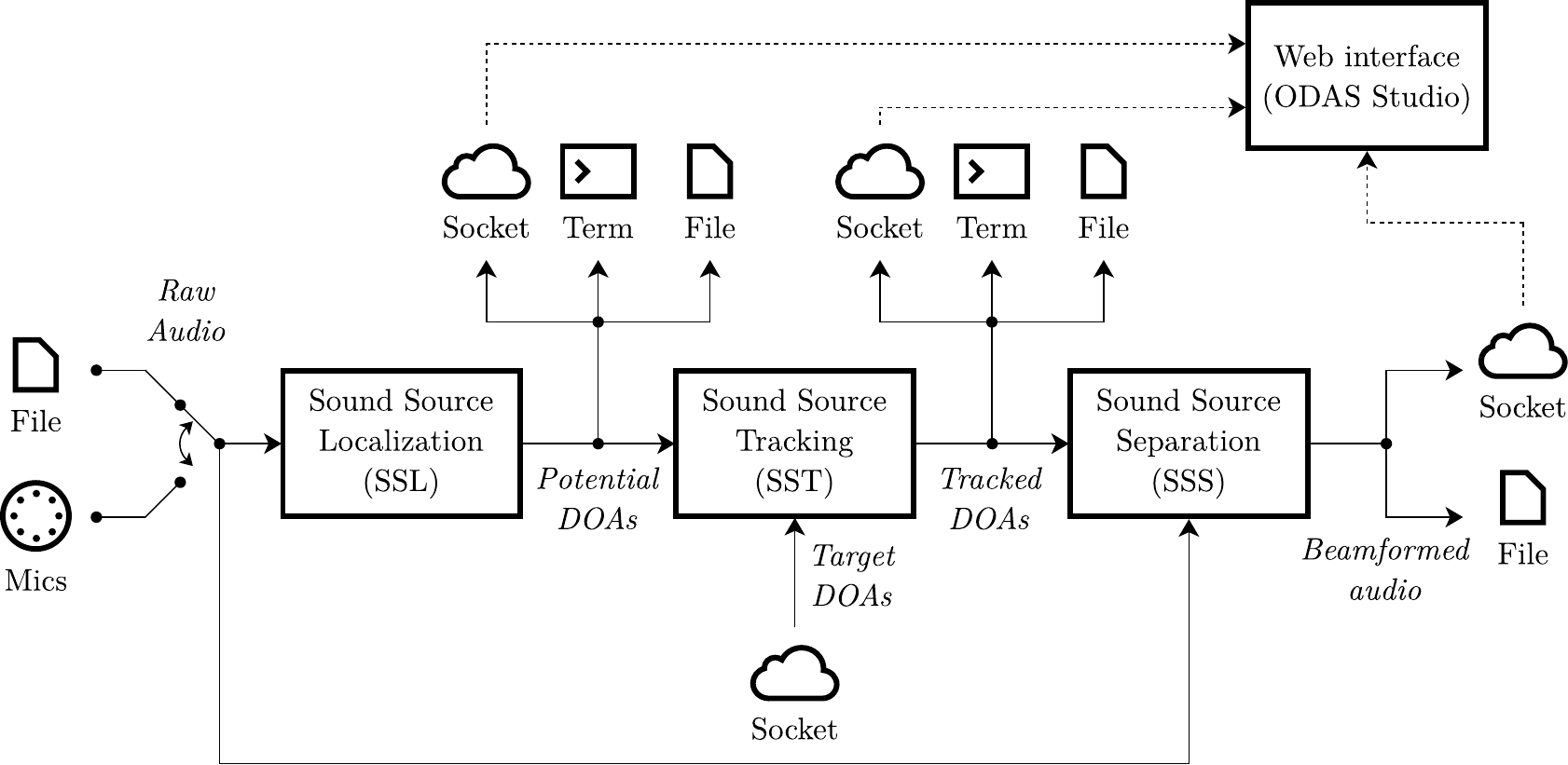}
    \caption{ODAS processing pipeline}
    \label{fig:pipeline}
\end{figure*}

Figure \ref{fig:pipeline} illustrates the audio pipeline and the I/O interfaces, each running in a separate thread to fully exploit processors with multiple cores.
Raw audio can be provided by a pre-recorded multi-channel RAW audio file, or obtained directly from a chosen sound card connected to microphones for real-time processing.
The Sound Souce Localization (SSL) module generates a fixed number of potential DOAs, which are fed to the Sound Source Tracking (SST) module.
SST identifies tracked sources, and these DOAs are used by the Sound Source Separation (SSS) module to perform beamforming on each target sound source.
DOAs can also be sent in JSON format to a terminal, to a file or to a TCP/IP socket.
The user can also define fixed target DOA(s) if the direction(s) of the sound source(s) is/are known in advance and no localization and no tracking is required.
The beamformed segments can be written in RAW audio files, or also sent via a socket.

\begin{figure}[!ht]
    \centering
    \includegraphics[width=\linewidth]{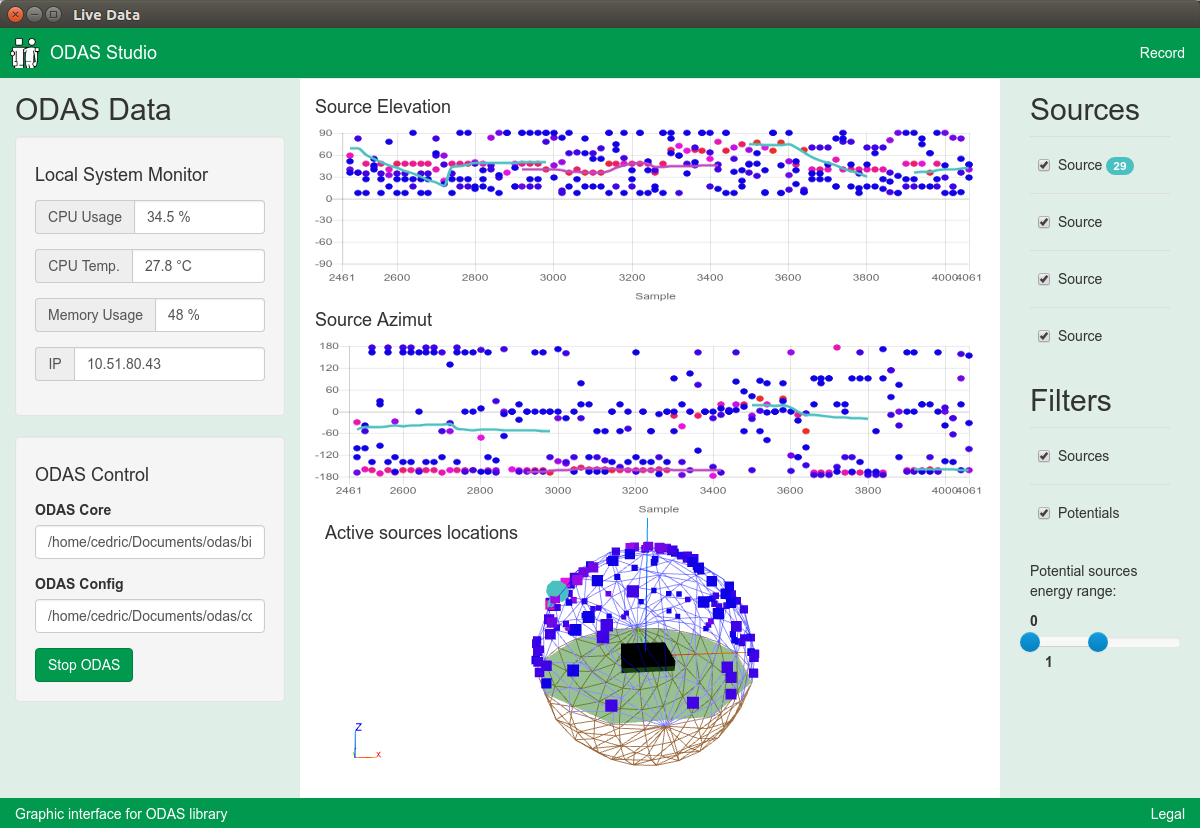}
    \caption{ODAS Studio Web Interface. Colored dots represent potential DOA with power levels, and solid lines illustrate the tracked sound source trajectories over time.}
    \label{fig:interface}
\end{figure}

The ODAS Studio Web Interface, shown in Fig. \ref{fig:interface}, can be used to visualize the potential and tracked DOAs, and also to get the beamformed audio streams.
This interface can run on a separate computer connected to ODAS via sockets.
The interface makes it possible to visualize the potential DOAs in three dimensions on a unit sphere with a color code that stands for their respective power, and in scatter plots of azimuths and elevations as a function of time.
The tracked sources are also displayed in the azimuth/elevation plots, as continuous lines with a unique color per tracked source.

ODAS relies on many strategies to reduce the computational load for the SSL, SST and SSS modules, described as follows.

\subsection{Sound Source Localization (SSL)}

\begin{figure}[!ht]
    \centering
    \includegraphics[width=0.8\linewidth]{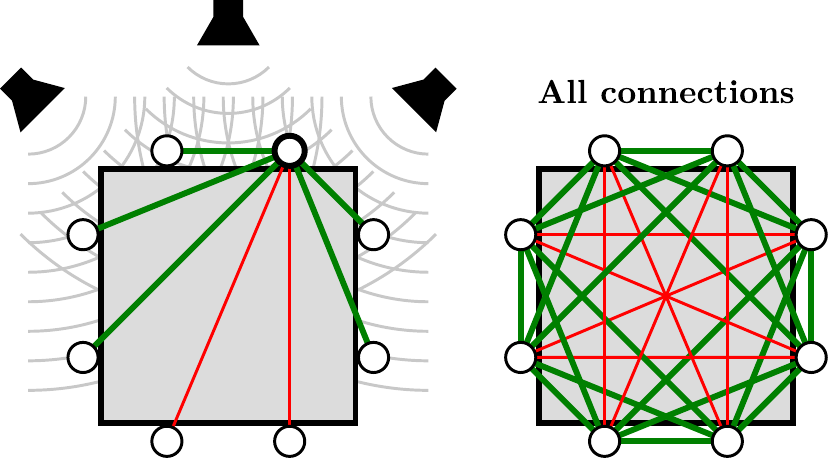}
    \caption{ODAS strategy exploiting microphone directivity to compute GCC-PHAT using relevant pairs of microphones in a closed array configuration}
    \label{fig:localisation_gccphat}
\end{figure}

ODAS exploits the microphone array geometry to perform localization, defined at start-up in a configuration file.
In addition to the position, the orientation of each microphone also provides useful information when microphones lie in a closed array configuration (e.g., when installed on a robot head or torso).
While microphones are usually omnidirectional, they can be partially hidden by some surfaces, which make their orientation relevant.
Localization relies on the Generalized Cross-Correlation with Phase Transform method (GCC-PHAT), computed for each pair of microphones.
ODAS uses the inverse Fast Fourier Transform (IFFT) to compute the cross-correlation efficiently from the signals in the frequency domain.
When dealing with small arrays, ODAS can also interpolate the cross-correlation signal to improve localization accuracy and to cope with the TDOA discretization artifact introduced by the IFFT.
While some approaches rely on the Head-Related Transfer Function (HRTF) to deal with closed array \cite{nakadai2003applying}, ODAS proposes a simpler model that provides accurate localization and reduce the computational load.
In fact, the framework exploits the directivity of microphones to only compute GCC-PHAT between pairs of microphones that can be simultaneously excited by the direct propagation path of a sound source.
To illustrate this, Fig. \ref{fig:localisation_gccphat} shows an 8-microphone closed array, for which it is assumed that all microphones point outside with a field of view of $180^\circ$.
Because microphones on opposite sides cannot capture simultaneously the sound wave coming from a source around the array, their cross-correlation can be ignored.
Consequently, ODAS computes GCC-PHAT for the pairs of microphones connected with green lines only. With such a closed array configuration, the pairs connected with red lines are ignored as there is no potential direct path for sound waves.
Therefore, ODAS computes the cross-correlation between $20$ pairs of microphones out of the $28$ possible pairs.
This simple strategy reduces the cross-correlation computational load by $29\%$ (i.e., $(28-20)/28$).
With open array configurations, ODAS uses all pairs of microphones because sound waves reach all microphones in such cases.

\begin{figure}[!ht]
    \centering
    \includegraphics[width=\linewidth]{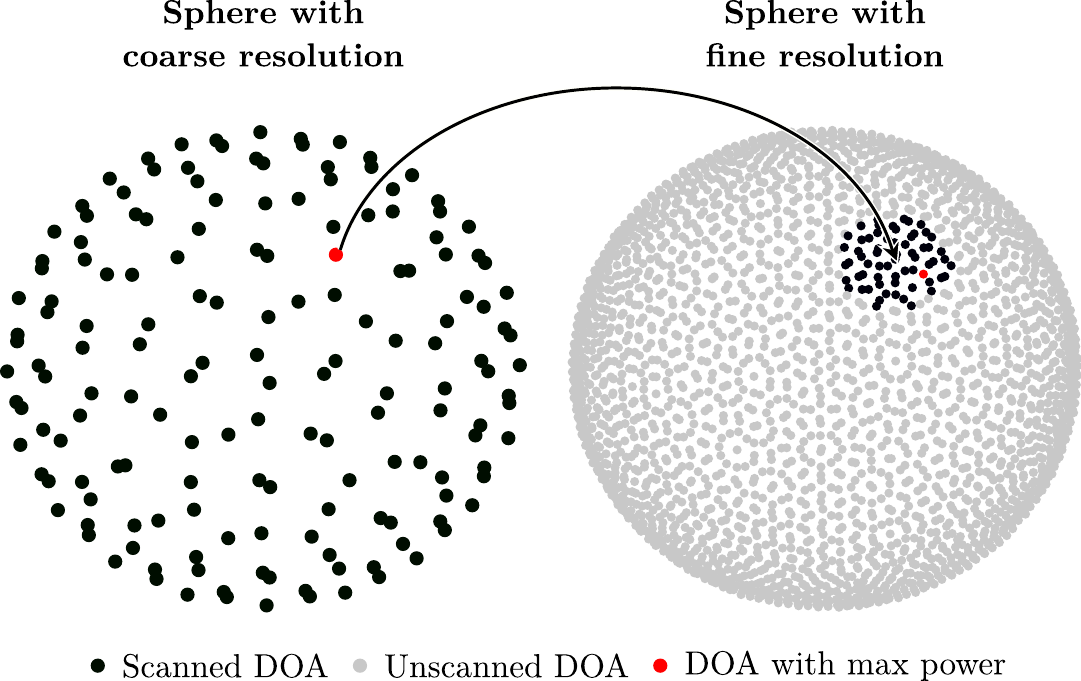}
    \caption{Illustration of the two unit sphere search, first with coarse resolution (left), and then more precise search with finer resolution (right)}
    \label{fig:sphere_coarse_fine}
\end{figure}

ManyEars computes the Steered-Response Power with Phase Transform (SRP-PHAT) for all DOAs that lie on a unit sphere discretized with $2562$ points.
For each DOA, ManyEars computes the SRP-PHAT power by summing the value of the cross-correlation associated to the corresponding time difference of arrival (TDOA) obtained with GCC-PHAT for each pair of microphones, and returns the DOA associated to the highest power.
Because there might be more than one active sound source at a time, the corresponding TDOAs are zeroed, and scanning is performed again to retrieve the next DOA with the highest power.
These successive scans are usually repeated to generate up to four potential DOAs.
However, scanning each point on the unit sphere involves numerous memory accesses that slow down processing.
To speed it up, ODAS uses instead two unit spheres: 1) one with a coarse resolution (made of $162$ discrete points) and 2) one with a finer resolution (made of $2562$ discrete points).
ODAS first scans all DOAs in the coarse sphere, finds the one associated to the maximum power, and then refines the search on a small region around this DOA on the fine sphere \cite{grondin2019lightweight}.
Figure \ref{fig:sphere_coarse_fine} illustrates this process, which reduces considerably the number of memory accesses while providing a similar DOA estimation accuracy.
For instance, when running ODAS on a Raspberry Pi 3, this strategy reduces CPU usage for performing localization using a single core for an 8-microphone array by almost a factor of 3 (from a single core usage of $38\%$ down to $14\%$) \cite{grondin2019lightweight}.
Note that when all microphones lie in the same plane in 2-D, ODAS scans only a half unit sphere, which also reduces the number of computations.

\subsection{Sound Source Tracking (SST)}

\begin{figure}
    \centering
    \includegraphics[width=\linewidth]{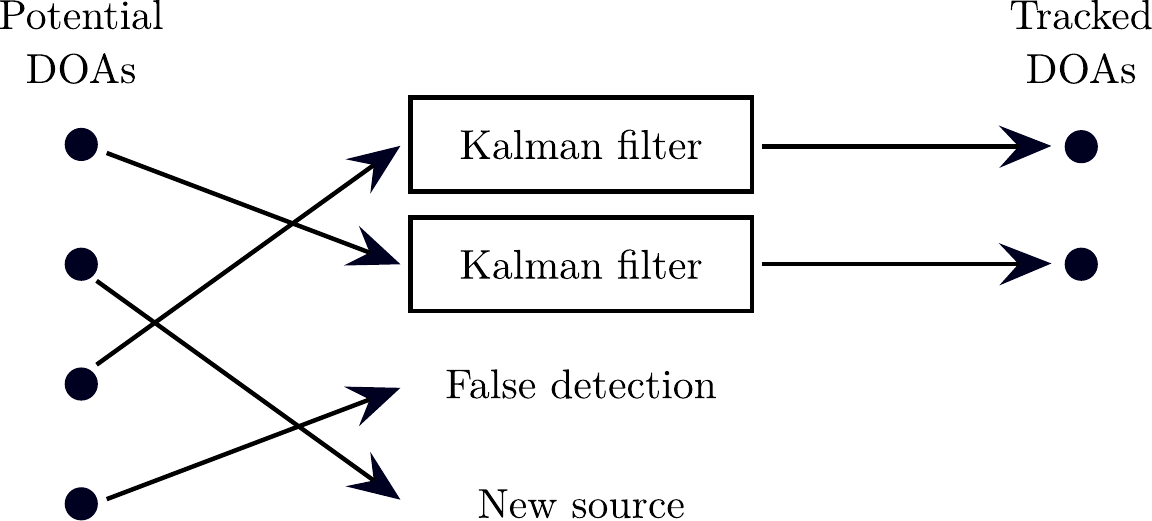}
    \caption{Tracking sound sources with Kalman filters: each DOA is associated to a previously tracked source, a false detection or a new source}
    \label{fig:tracking}
\end{figure}

Sound sources are non-stationary and sporadically active over time.
Tracking therefore provides a continuous trajectory in time for the DOA of each sound source, and can cope with short silence periods.
This module also detects newly active sound sources, and forgets sources that are inactive for a long period of time.
Sound source localization provides one or many potential DOAs, and the tracking maps each observation either to a previously tracked source, to a new source, or to a false detection.
To deal with static and moving sound sources, ManyEars also relies on a particle filter to model the dynamics of each source \cite{grondin2013manyears}.
The particles of each filter are associated to three possible states: 1) static position, 2) moving with a constant velocity, 3) accelerating.
This approach however involves a significant amount of computations, as the filter is usually made of $1000$ particles, and each of them needs to be individually updated.
\cite{briere2008embedded} proposed to reduce the number of particles by half to bring down the computational load. Experiments however demonstrated that this impacts accuracy when tracking two moving sound sources.
Instead ODAS uses Kalman filter for each tracked source, as illustrated by Fig. \ref{fig:tracking}.
Results demonstrate similar tracking performance, with a significant reduction in computational load on a Raspberry Pi 3 device (e.g., by a factor of $30$, from a single core usage of $24\%$ down to $0.8\%$ when tracking one source, and by a factor of $14$, from a single core usage of $98\%$ down to $7\%$ when tracking four sources \cite{grondin2019lightweight}).

\subsection{Sound Source Separation (SSS)}

\begin{figure}[!ht]
    \centering
    \includegraphics[width=0.6\linewidth]{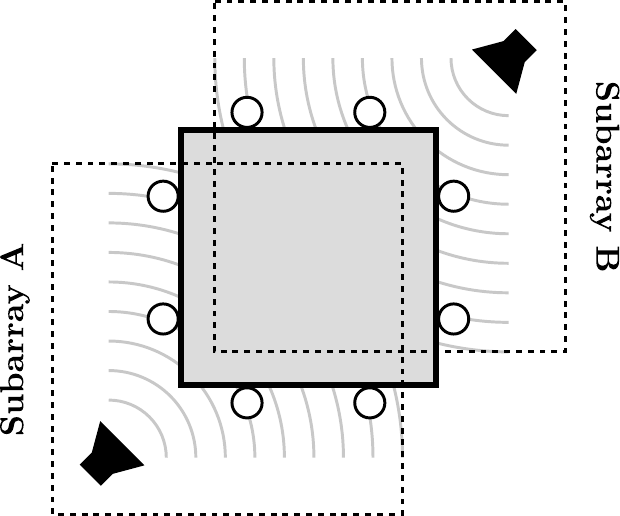}
    \caption{ODAS subarray SSS strategy to optimize processing}
    \label{fig:separation}
\end{figure}

ODAS supports two sound source separation methods: 1) delay-and-sum beamforming, and 2) geometric sound source separation.
These methods are similar to the former methods implemented in ManyEars \cite{grondin2013manyears}, with the exception that ODAS also considers the orientation of each microphone.
Because ODAS estimates the DOA of each sound source, it can select only the microphones oriented in the direction of the target source for beamforming.
For a closed array configuration, this improves separation performance (e.g., when using a delay-and-sum beamformer, this can result in a SNR increase of 1 dB when compared to using all microphones \cite{grondin2017systeme}), while it reduces the amount of computations.
Figure \ref{fig:separation} presents an example with two sound sources around a closed array, where ODAS performs beamforming with the microphones on the left and bottom to retrieve the signal from the source on the left, and performs beamforming with the microphones on the right and top to retrieve the signal from the source on the right.
Subarray A only uses four out of the eight microphone, and subarray B uses the other four microphones. 
This reduces the amount of computations and also improves separation performance.

ODAS also implements the post-filtering approach formerly introduced in ManyEars \cite{grondin2013manyears}.
Post-filtering aims to improve speech intelligibility by masking time-frequency components dominated by noise and/or competing sound source \cite{valin2004enhanced}.

ODAS also implements the post-filtering approach formerly introduced in ManyEars \cite{grondin2013manyears}.
Post-filtering aims to improve speech intelligibility by masking time-frequency components dominated by noise and/or competing sound source \cite{valin2004enhanced}.

\section{Configuring the ODAS library}
\label{sec:lib}

ODAS relies on a custom configuration file that holds all parameters to instantiate the modules in the processing pipeline, with some parameters determined by the microphone array hardware.
There are some useful guidelines to follow when the microphone array geometry can be customized: 1) the microphones should span all x, y and z dimensions to localize sound sources in the full elevation and azimuth ranges. When microphones only span two dimensions, the localization is limited to a half sphere; 2) the microphones should be a few tens of centimeters apart. Increasing the microphone array aperture provides better discrimination in low-frequencies, which is well-suited for speech; 3) using a directivity model for closed arrays with microphones installed on rigid surface further reduces the computational load and improves accuracy.
The structure of each file obeys the configuration format, which is compact and easy to read\footnote{\url{http://hyperrealm.github.io/libconfig/}}.
The configuration file is divided in many sections:

\subsubsection{Raw input}

This section indicates the format of the RAW audio samples provided by the sound card or the pre-recorded file.
It includes the sample rate, the number of bits per sample (assuming signed numbers), the number of channels and the buffer size (in samples) for reading the audio.

\subsubsection{Mapping}

The mapping selects which channels are used as inputs to ODAS.
In fact, some sound cards have additional channels (e.g., for playback) and it is therefore convenient to extract only the meaningful channels.
Moreover, this option allows a user to ignore some microphones if desired to reduce computational load.

\subsubsection{General}

This section provides general parameters that are used by all the modules in ODAS' pipeline.
It includes the short-time Fourier Transform (STFT) frame size and hop length (since all processing is performed in the frequency domain).
It also provides a processing sample rate, which can differ from the initial sample rate from RAW audio in the sound card (ODAS can resample the RAW signal to match the processing sample rate).
The speed of sound is also provided, along with some uncertainty parameter, to cope with different speeds of sound.
All microphone positions are also defined, along with their orientation.
It is also possible to incorporate position uncertainty to make localization more robust to measurement errors when dealing with microphone arrays of arbitrary shape.

\subsubsection{Stationary Noise Estimation}

ODAS estimates the background noise using the minima controlled recursive averaging (MCRA) method \cite{cohen2002noise}, to make localization more robust and increase post-processing performances.
This section provides the MCRA parameters to be used by ODAS.

\subsubsection{Localization}

The localization section provides parameters to fine tune the SSL module.
These parameters are usually the same for all setups, except for the interpolation rate which can be increased when dealing with small microphone arrays to cope with the discretization artifact introduced by the IFFT when computing GCC-PHAT.

\subsubsection{Tracking}

ODAS can support tracking with either the former particle filter method, or the current Kalman filter approach.
Most parameters in this section relate to the methods introduced in \cite{grondin2019lightweight}.
It is worth mentioning that ODAS represents the power distribution for a DOA generated by the SSL module as a Gaussian Mixture Model (GMM).
Another GMM also models the power of diffuse noise when all sound sources are inactive.
It is therefore possible to measure both distributions experimentally using histograms and then fit them with GMMs.

\subsubsection{Separation}

This section of the configuration file defines ODAS which separation method to use (Delay-and-sum or Geometric Source Separation).
It also provides parameters to perform post-filtering, and information regarding the audio format of the separated and post-filtered sound sources.

A configuration file can be provided for each commercial microphone array or each robot with a unique microphone array geometry, and to use ODAS for processing.

\section{Applications}
\label{sec:applications}

\begin{figure}
    \centering
    \begin{subfigure}[b]{0.49\linewidth}
        \includegraphics[width=\linewidth]{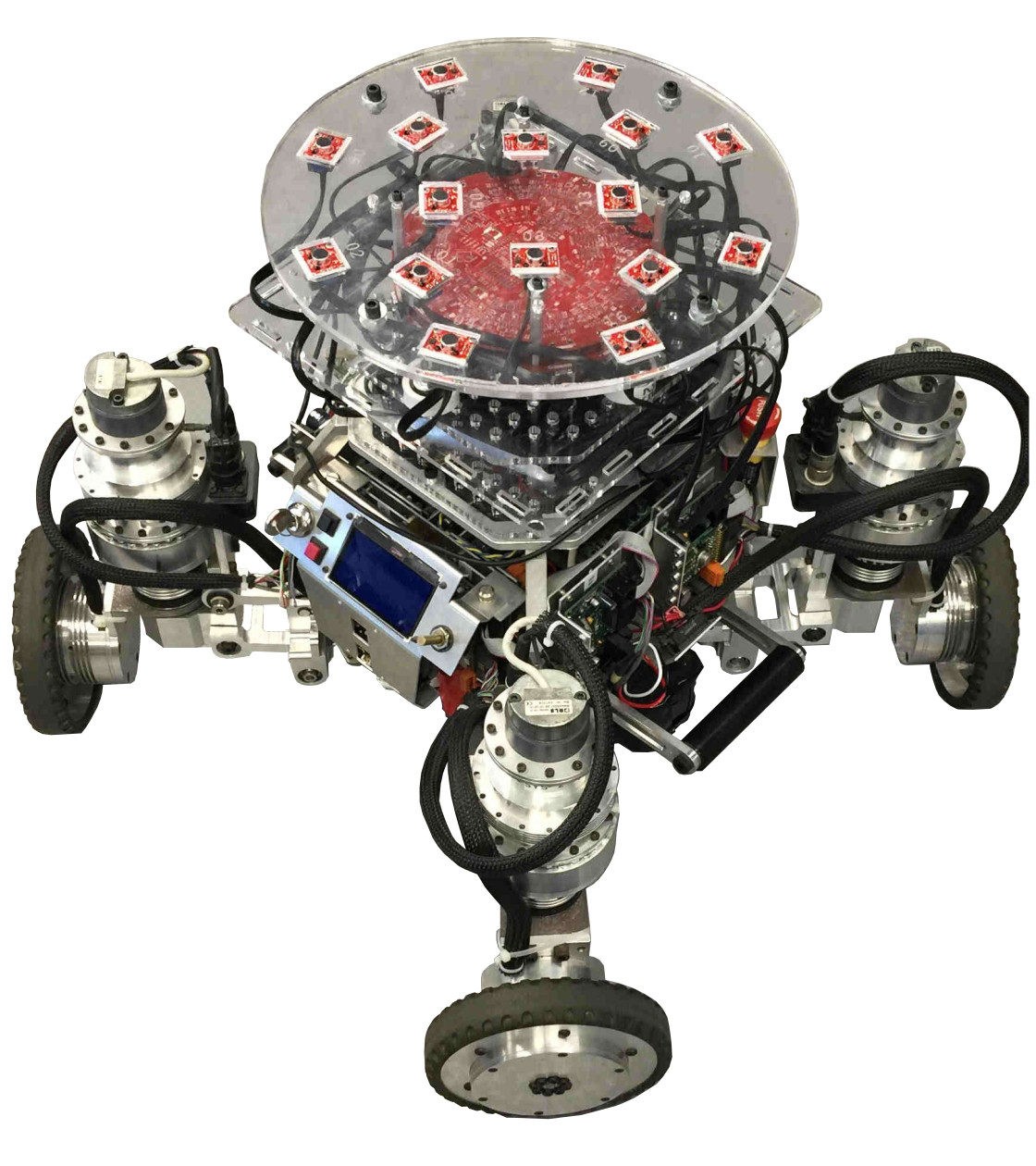}
        \subcaption{Azimut-3 (open)}
        \label{fig:azimut3_open}
    \end{subfigure}
    \begin{subfigure}[b]{0.49\linewidth}
        \includegraphics[width=\linewidth]{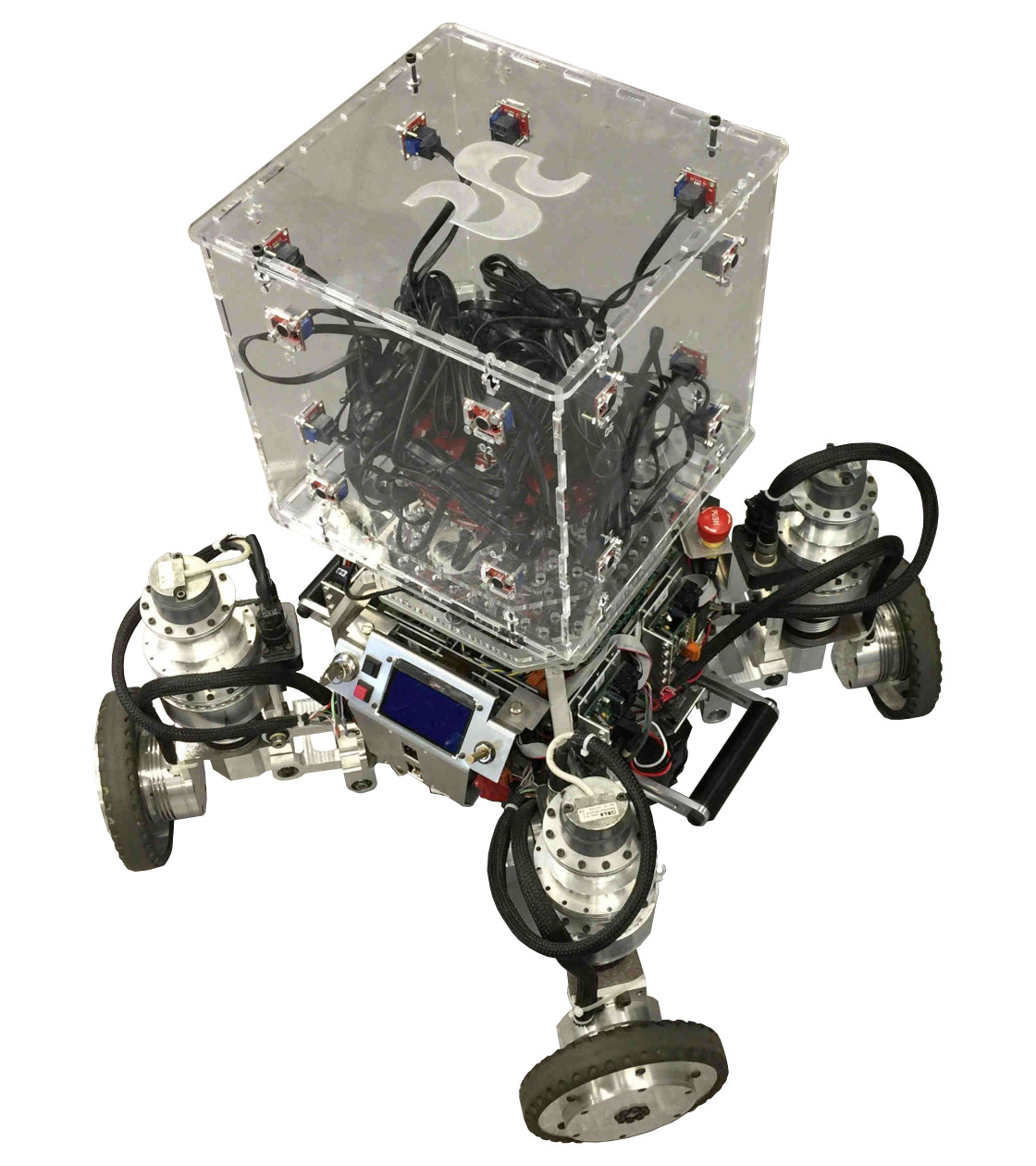}
        \subcaption{Azimut-3 (closed)}
        \label{fig:azimut3_closed}
    \end{subfigure}\\
    \medskip
    \begin{subfigure}[b]{0.49\linewidth}
        \includegraphics[width=\linewidth]{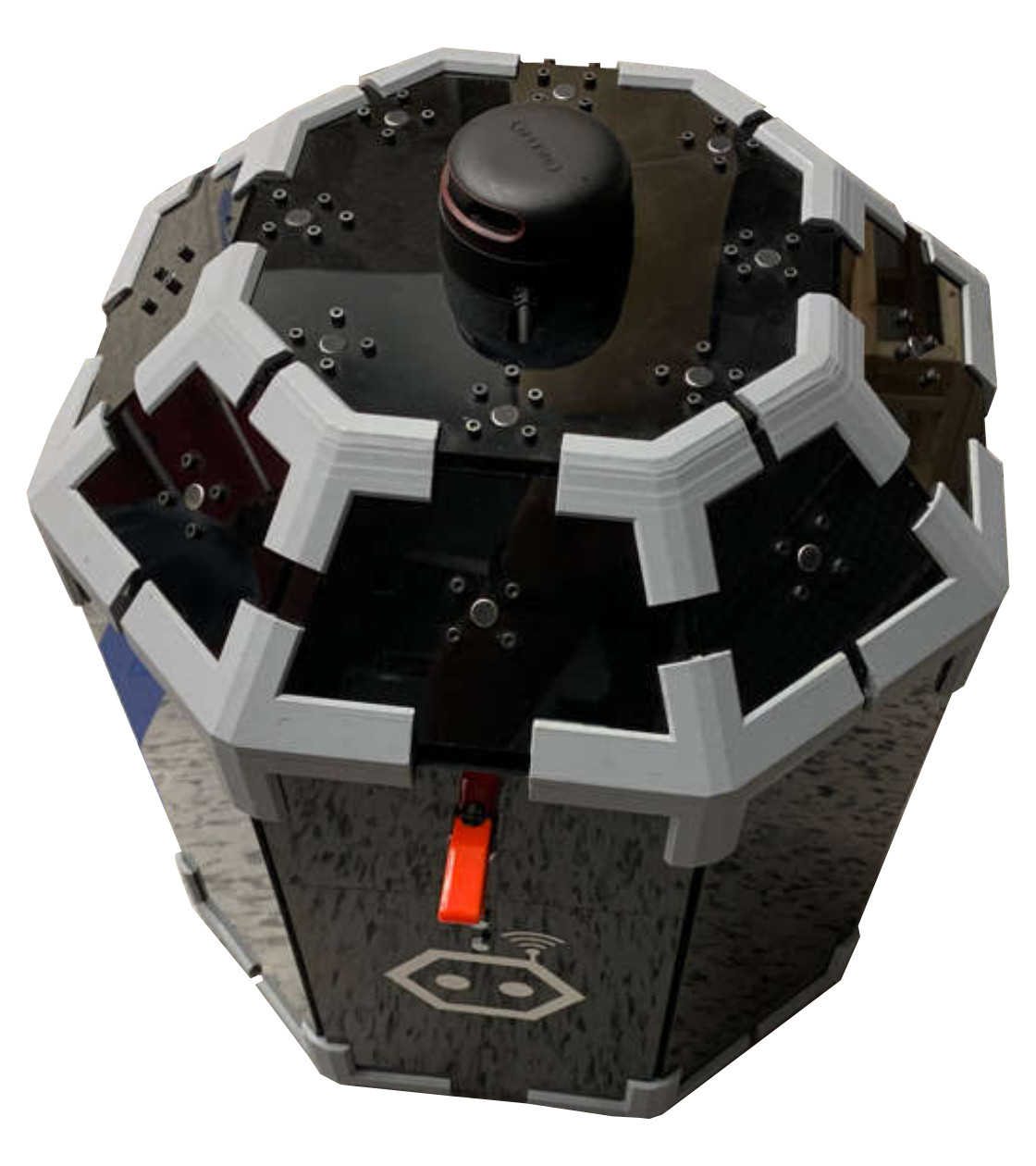}
        \subcaption{SecurBot}
        \label{fig:securbot}
    \end{subfigure}
    \begin{subfigure}[b]{0.49\linewidth}
        \includegraphics[width=\linewidth]{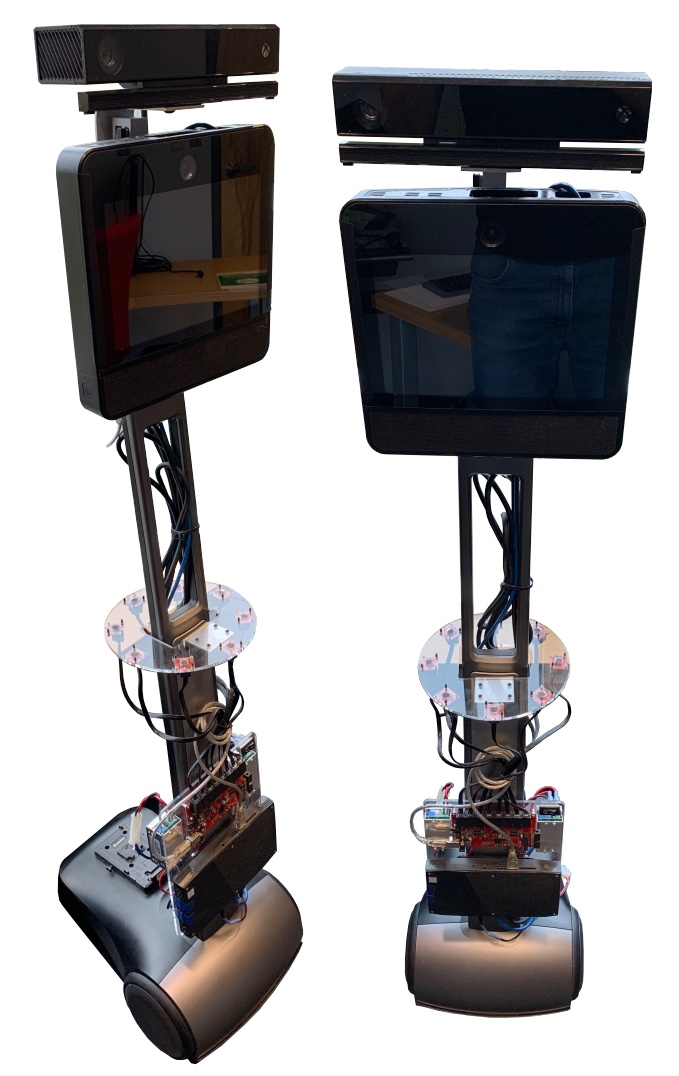}
        \subcaption{Beam}
        \label{fig:beam}
    \end{subfigure}
    \caption{ODAS with the Azimut-3 (open and closed array configurations, 16 microphones), SecurBot (16-microphone configuration on top and sides) and Beam (8-microphone on a circular support) robots}
    \label{fig:robots}
\end{figure}

ODAS' optimized processing strategies makes it possible to perform all processing on low-cost hardware, such as a Raspberry Pi 3 board.
Figures \ref{fig:azimut3_open}, \ref{fig:azimut3_closed}, \ref{fig:securbot} and \ref{fig:beam} present some of the robots using the ODAS framework for sound source localization, tracking and separation.
ODAS is used with the Azimut-3 robot, with two different microphone array configurations \cite{grondin2019lightweight}: 16 microphones lying on a single plane, or with all microphones installed around a cubic shape on top of the robot.
For both setups, the sound card 16SoundsUSB\footnote{\url{https://github.com/introlab/16SoundsUSB}} performs signal acquisitions and then ODAS performs SSL, SST and SSS.
Figure \ref{fig:securbot} illustrates the 16 microphones configuration  on the SecurBot robots \cite{michaud20203d}.
ODAS is also used with the Beam robot \cite{laniel2017adding}, placing 8 microphones on the same plane and using the 8SoundsUSB sound card\footnote{\url{https://sourceforge.net/p/eightsoundsusb/wiki/Main_Page/}}.
ODAS exploits the directivity of microphones for setups with the Azimut-3 robot (closed array configuration) and SecurBot, which reduces the amount of computations.
On the other hand, ODAS searches for DOAs on a half sphere for the Azimut-3 robot (open array configuration) and the Beam robot, as all microphones lie on the same 2-D plane.
The proposed framework is also used with the T-Top robot \cite{maheux2022t}, which base is equipped with 16 microphones.
The Open-Tera \cite{panchea2022opentera} microservice architecture solution for telepresence robots also uses the ODAS framework.

ODAS is also used for drone localization using three static 8-microphone arrays disposed on the ground at the vertex of an equilateral triangle, with edges of 10 m \cite{lauzon2017localization}.
DOA search is performed on a half-sphere over the microphone array plane.
Each microphone array is connected to a Raspberry Pi 3 through the 8SoundsUSB sound card, which runs ODAS and returns potential DOAs to a central node.
The central node then performs triangulation to estimate the 3D position of the flying drone.

ODAS is also extensively used with smaller commercial microphone arrays, for sound source localization or as a preprocessing step prior to speech recognition and other recognition tasks.
These are usually small circular microphone arrays, where the number of microphones varies between $4$ and $8$.
ODAS is referenced on Seeed Studio ReSpeaker USB 4-Mic Array official website as a compatible framework with their hardware\footnote{\url{https://respeaker.io/4\_mic\_array/}}. 
The Matrix Creator board has numerous sensors, including eight microphones on its perimeters, and has online tutorials showing how to use ODAS with the array\footnote{\url{https://www.youtube.com/watch?v=6ZkZYmLA4xw}}.
ODAS was also validated with the miniDSP UMA-8 microphone array\footnote{\url{https://www.minidsp.com/aboutus/newsletter/\\listid-1/mailid-68-minidsp-newsletter-an-\\exciting-new-chapter}}, and the XMOS xCore 7-microphone array\footnote{\url{https://www.youtube.com/watch?v=n7y2rLAnd5I}}.
For all circular arrays, ODAS searches for DOAs on a half-sphere, and also interpolates the cross-correlation results to improve accuracy since microphones are only a few centimeters apart.

Configuration files with the exact positions of all microphones for each device are available online with the source code.

\section{Conclusion}

This paper introduces ODAS, the Open embeddeD Audition System framework, explaining its strategies for real-time and embedded processing, and demonstrates how it can be used for various applications, including robot audition, drone localization and voice assistants.
ODAS' strategies to reduce computational load, consist of: 1) partial cross-correlation computations using the microphone directivity model, 2) DOA search on coarse and fine unit spheres, 3) search on a half sphere when all microphones lie on the same 2-D plane, 4) tracking active sound sources with Kalman filters and 5) beamforming with subarrays using simple microphone directivity models.
In addition to use cases found in the literature, ODAS source code has been accessed more than 55,000 times, which suggests that there is a need for a framework for robot audition that can run on embedded computing systems.
ODAS can also be part of the solution for edge-computing for voice recognition to avoid cloud computing and preserve privacy.

While ODAS performs well in quiet or moderately noisy environments, its robustness should be improved for more challenging environments (noisy conditions, strong reflections, important ego-noise, etc.). In future work, additional functionalities will be added to ODAS, including new algorithms that rely on deep learning based methods, as machine learning has became a powerful tool when combined with digital signal processing for sound source localization \cite{chakrabarty2019multi}, speech enhancement \cite{valin2018hybrid,valin2020perceptually} and sound source classification \cite{ford2019deep,grondin2019sound}.
Additional beamforming methods could also be implemented, including Minimum Variance Distortionless Response (MVDR) beamformer \cite{habets2009new}, and generalized eigenvalue (GEV) beamforming \cite{heymann2015blstm,grondin2020gev}, as these approaches are particularly suited for preprocessing before automatic speech recognition.
ODAS would also benefit from ego-noise suppression algorithms to mitigate the impact of motor noise while doing sound source localization, tracking and separation.

\section*{Conflict of Interest Statement}

The authors declare that the research was conducted in the absence of any commercial or financial relationships that could be construed as a potential conflict of interest.

\section*{Author Contributions}

FG designed and wrote the ODAS framework C library. DL assisted with maintenance and integration of the library on all robotics platforms. CG designed the ODAS studio web interface. JSL and JV debugged the ODAS library. SM and SF worked on integrating the framework in ROS. FM supervised and led the team.

\section*{Funding}
This work was supported by FRQNT -- Fonds recherche Qu\'ebec Nature et Technologie.

\bibliography{bibliography}
\bibliographystyle{IEEEtran}

\end{document}